\documentclass[12pt]{article}
\usepackage{graphicx}
\begin{document}
\baselineskip=5.5mm
\newcommand{\be} {\begin{equation}}
\newcommand{\ee} {\end{equation}}
\newcommand{\Be} {\begin{eqnarray}}
\newcommand{\Ee} {\end{eqnarray}}
\def\lg{\langle}
\def\rg{\rangle}
\def\a{\alpha}
\def\b{\beta}
\def\g{\gamma}
\def\G{\Gamma}
\def\d{\delta}
\def\D{\Delta}
\def\e{\epsilon}
\def\k{\kappa}
\def\l{\lambda}
\def\L{\Lambda}
\def\om{\omega}
\def\Om{\Omega}
\def\t{\tau}
\noindent
\noindent
\begin{center}
{\Large
{\bf
Nonresonant holeburning in the Terahertz range: Brownian oscillator model
}}\\
\vspace{0.8cm}
{\bf Uli H\"aberle and Gregor Diezemann} \\
Institut f\"ur Physikalische Chemie, Universit\"at Mainz,
Welderweg 11, 55099 Mainz, FRG
\\
\end{center}
\vspace{1cm}
\noindent
{\it
The response to the field sequence of nonresonant hole burning, a
pump-wait-probe experiment originally designed to investigate slow relaxation
in complex systems, is calculated for a model of Brownian oscillators,
thus including inertial effects.
In the overdamped regime the model predictions are very similar to those
of the purely dissipative stochastic models investigated earlier,
including the possibility to discriminate between dynamic homogeneous and
heterogeneous relaxation.
The case of underdamped oscillations is of particular interest when
low-frequency excitations in glassy systems are considered.
We show that also in this situation a frequency selective modification of
the response should be feasable.
This means that it is possible to specifically address various parts of the
spectrum.
An experimental realization of nonresonant holeburning in the Terahertz regime
therefore is expected to shed further light on the nature of the vibrations
around the so-called boson peak.
}

\vspace{0.5cm}

\vspace{1cm}
\noindent
\section*{I. Introduction}
The relaxation functions observed in disordered materials such as glasses,
spin-glasses, disordered crystals or proteins are usually found to decay
non-exponentially on macroscopic time scales\cite{het.dyn}.
Several experimental techniques have been invented in order to investigate the
detailed nature of this non-exponentiality, among them a reduced
four-dimensional NMR technique\cite{SRS91}, an optical deep bleach
experiment\cite{CE95} and also nonresonant holeburning (NHB)\cite{nhb.sci}.
Common to all the techniques is that they allow to specifically select a
(slow) sub-ensemble and afterwards monitor its relaxation.
This way the existence of {\it dynamic} heterogeneities could be
verified experimentally\cite{4dnmr}.
In the present paper we adopt the definition given in ref.\cite{het.vigo}
according to which a response or relaxation function is termed dynamic
heterogeneous if it is possible to specifically address effectively slow,
intermediate or fast contributions to the ensemble averaged function.
Otherwise, the relaxation will be termed dynamic homogeneous.
It is important to mention that neither this definition nor any of the quoted
experiments allow to obtain information about any spatial aspects of these
heterogeneities.

NHB consists of a pump-wait-probe field sequence, starting with the
application of one (or more) cycles of a large amplitude ac field with
frequency $\Om$ to the sample in equilibrium.
After a waiting time $t_w$ has elapsed, a small field is turned on to
monitor the modified response, cf. Fig.1.
A qualitative interpretation of the experimental results relies on the fact
that via the application of the ac field the sample absorbs energy of an
amount proportional to the imaginary part of the susceptibility evaluated
at the pump-frequency $\Om$\cite{kubo}.
One thus expects that a frequency selective modification of the spectrum should
be feasible only if the response is given by a heterogeneous superposition of
entities relaxing at different rates.
This view has proven fruitful in the interpretation of experimental results
obtained from a variety of samples, including supercooled
liquids\cite{nhb.sci, ranko03}, disordered crystals\cite{KSB98}, amorphous
ion-conductors\cite{RB99} and spin glasses\cite{ralph99}.
A theoretical investigation in terms of a response theory has been
developed for the case of stochastic dipole-reorientations\cite{epl.nhb},
which has shown that the modified response indeed depends on the
absorbed energy, thus supporting the above picture\cite{BD02}.

At this point a note of caution is appropriate. NHB differs fundamentally from
the so-called spectral holeburning experiment known from nonlinear
optics\cite{mukamel}, which e.g. has successfully been used to investigate the
dynamics of two-level systems in glasses\cite{friedrich}.
In this experiment the optical transition of a dye molecule is altered
externally, thus giving rise to a hole in the remaining absorption spectrum.
In contrast, NHB does not work at frequencies allowing to monitor electronic
transitions and we are concerned with dynamic features which are much slower.

Because the mentioned theoretical investigation on NHB\cite{epl.nhb} was
concerned with slow reorientational dynamics, typically on a time scale of
$\mu s\cdots s$, only purely dissipative dynamics has been considered and
inertial effects have been neglected completely.
Given the fact that the primary relaxation of supercooled liquids is of a
dynamic heterogeneous nature, the question as to which extent the same holds
for the fast dynamics in these systems naturally arises.
In particular, this point may prove of importance for understanding
the physical origin of the so-called boson peak, observable e.g. as
low-frequency oscillations in nonresonant Raman scattering\cite{boson}.
On the experimental side, the recent developments in far-infrared
spectroscopy\cite{BTS02}, in particular the application of femtosecond
Terahertz pulses\cite{ZHD90} have opened the way to investigate the nonlinear
dielectric response in this frequency range in detail.
Thus, there is good reason to hope that the realization of an experimental
protocol like the NHB field sequence should not be out of reach in due
course\cite{keith.exp}.

In this paper we present a theoretical analysis of NHB for the case of
damped oscillations using the well known model of Brownian oscillators
(BO)\cite{mukamel}.
This model, which often is employed in calculations of the nonlinear optical
response in condensed phases, allows us to treat both, underdamped and
overdamped motions in a coherent way.
The outline of the paper is as follows. In the next section we describe the
model and the theoretical aspects of our study.
Section III contains the results of the model calculations along with the
discussion, including an estimate of the expected magnitude of the nonlinear
effects. Finally, we close with some concluding remarks in Sect. IV.
\section*{II. NHB for the Brownian Oscillator Model}
In this section we briefly discuss those features of the BO-model that will
be relevant for our discussion.
Additionally, we calculate the linear and the third-order response functions
relevant for the NHB experiment.
As we are primarily concerned with the response of amorphous systems, we have
to face the problem of treating spatial correlations of the coefficients in
a normal-mode expansion of the dipole moments. Therefore a discussion of the
frequency dependence of the so-called light to vibration coupling constant is
included.
\subsection*{A. General aspects and linear response}
We assume that the normal-modes $\{q\}=\{q_1,q_2,\cdots\}$ of the system under
consideration in the THz regime can be approximated as independent oscillators.
In a molecular system, the $q_i$ would correspond to internal degrees of
freedom like vibrational or torsional modes and in a macroscopic system like
an amorphous system the $q_i$ represent all vibrational excitations including
the low-frequency phonon-like collective modes.

In the classical version of the BO-model with Markovian damping (Ohmic
friction)\cite{mukamel} and equal masses, $m_i=m$, the dynamics of the $q_i$
is described by the Langevin equation:
\be\label{Lang.gl}
m \ddot{q}_i+m \g_i \dot{q}_i+\frac{d V(\{q\})}{d q_i}=\G_i(t)
-{\partial {\cal H}_{int} \over \partial q_i}
\ee
Here, $\g_i$ denotes the damping constant and $\G_i(t)$ is a Gaussian
stochastic force with zero mean and correlation
$\lg \G_i(t)\G_k(t')\rg =2\d_{i,k}\g_i \b^{-1} m\d(t-t')$, $\b^{-1}=k_BT$.
Furthermore, the coupling to the field, ${\cal H}_{int}$, is given
by\cite{mukamel}:
\be\label{H.int}
{\cal H}_{int}=-\int\! d{\bf r}\mu({\bf r},\{q\})E({\bf r},t).
\ee
The Fokker-Planck equation corresponding to the Langevin equation,
eq.(\ref{Lang.gl}), is solved using the harmonic case as the unperturbed model
and treating anharmonic terms in $V(\{q\})$ as well as ${\cal H}_{int}$ in
perturbation theory similar to the calculation performed in ref.\cite{epl.nhb}.
We mention that the assumption of Ohmic friction can easily be relaxed and a
dynamic homogeneous scenario can be modelled via a frequency dependent damping
$\g(\om)$. This case, however, is most relevant for overdamped motions and
will be treated in a forthcoming publication\cite{uli.ou}.
The solution of eq.(\ref{Lang.gl}) shows that all quantities of physical
interest are determined by $\l_{i,1}={1\over 2}(\g_i+\d_i)$ and
$\l_{i,2}={1\over 2}(\g_i-\d_i)$ with
$\d_i=\left(\g_i^2-4\om_i^2\right)^{1/2}$.

In order to calculate the response to an external electric field, we expand
the dipole moment
\Be\label{mu.q}
\mu({\bf r},\{q\}) = \mu_0({\bf r})
+\sum_i\mu({\bf r})'_i q_i+{1\over 2}\sum_i\mu({\bf r})''_i q_i^2+\cdots
\nonumber\\
\mbox{with}\quad
\mu({\bf r})'_i=\left({\partial \mu({\bf r})\over \partial q_i}\right)_0
\quad\mbox{and}\quad
\mu({\bf r})''_i=\left({\partial^2 \mu({\bf r})\over \partial q_i^2}\right)_0
\Ee
and neglect higher order terms as well as cross-terms for simplicity.
We will focus on the dielectric response but the inclusion of a similar
expansion of the polarizability, relevant for Raman scattering, poses
no problem.
Note that we have kept the position-dependence of the dipole moment, since
in amorphous systems correlations are expected to exist which extent over
some finite volume, depending on the coherence length of the modes
$q_i$\cite{SG70}.

In the present model, there are different contributions to a non-vanishing
third-order response\cite{OT98}. One nonlinear term stems from the quadratic
term in eq.(\ref{mu.q}).
Another possible source of nonlinearity has its origin in anharmonic
contributions to the potential.
In our calculations, we use a potential of the form $V(\{q\})=\sum_i V(q_i)$
with
\be\label{V.qi}
V(q_i)=m\om_i^2\left[{1\over 2}q_i^2+{1\over 3}\Theta_3q_i^3+
{1\over 4}\Theta_4q_i^4\right]+{\cal O}(q_i^5)
\ee
where we have scaled the anharmonicity strengths by the harmonic frequency
$\om_i^2$. Higher-order terms in the potential and also a coupling among
the $q_i$ can easily be incorporated albeit this leads to an increasing number
of parameters.

The polarization following the NHB field sequence for a single mode $q_i$
is calculated in ${\cal O}(E_P^2)$ and ${\cal O}(E_M)$.
Without going into the details of the calculations here, we mention that
the NHB-response $P_i^{*(X)}(t,t_w,t_p)$ is given as a superposition of the
ordinary linear response and a modification:
\be\label{Pi.star}
P_i^{*(X)}(t,t_w,t_p)=P_i^{\rm(X)}(t)+\D P_i^{\rm(X)}(t,t_w,t_p)
\ee
Here, $P_i^{\rm(X)}(t)$ denotes the polarization in ${\cal O}(E_M)$.
In addition, the nonlinear modification, $\D P_i^{\rm(X)}(t,t_w,t_p)$, which
is ${\cal O}(E_P^2\cdot E_M)$, contains various terms, to be discussed later.
Note that the form of the signal is formally identical to the one found in
ref.\cite{epl.nhb}. Thus the same phase-cycling as described in
ref.\cite{nhb.sci} can be applied in order to extract the modification
$\D P_i^{\rm(X)}$.
The superscripts 'X' stand for the different response functions measured.
In the present paper we will discuss three different functions.
A field pulse yields the pulse-response (X=P), while application of a constant
field gives the step-response (X=S), also called the integrated response.
Furthermore, we consider the response to an oscillatory field,
denoted by X=AC.
Summarizing, we have:
\[
\mbox{X=S}:E_M(t)=E_M\theta(t)
\quad;\quad\mbox{X=P}:E_M(t)=E_M\d(t)
\quad ; \quad\mbox{X=AC}:E_M(t)=E_Me^{-i\om t}
\]
where the fields in eq.(\ref{H.int}) are given by
$E({\bf r},t)=e^{i{\bf k}{\bf r}}E(t)$.

Denoting the wave vector of the measuring field by ${\bf k}_M$,
one has for the polarization in the linear response regime:
\be\label{Pi.lin}
P_i^{\rm(X)}(t)={\rho\over m}E_M\int\!d{\bf r}_0\int\!d{\bf r}_1
			\mu({\bf r}_0)_i'\mu({\bf r}_1)_i'
			e^{i{\bf k}_M{\bf r}_1}R_i^{\rm(X)}(t)
\ee
Here, $\rho=N/V$ is the number density of oscillators and $R_i^{\rm(X)}(t)$
denotes the linear response of mode '$i$'. The latter are different for
different time-dependencies of $E_M(t)$.

In case of the pulse-response we have:
\be\label{Ri.p.t}
R_i^{\rm(P)}(t)={1\over\d_i}\left(e^{-\l_{i,2}t}-e^{-\l_{i,1}t}\right)
\ee
The integrated response function is trivially related to the pulse-response
via $R_i^{\rm(S)}(t)=\int_0^t\!d\t R_i^{\rm(P)}(\t)$
and the response to an ac field of frequency $\om$ reads as:
\be\label{Ri.ac.t}
R_i^{\rm(AC)}(t)={1\over\d_i}\left(\frac{1}{\l_{i,2}-i\om}
					(e^{-i \om t}-e^{-\l_{i,2}t})
				  -\frac{1}{\l_{i,1}-i\om}
					(e^{-i \om t}-e^{-\l_{i,1}t})
			     \right)
\ee
In the steady state, characterized by $e^{-\l_{i,1/2}t}\!=\!0$, this expression
coincides with the fourier transform of $R_i^{\rm(P)}(t)$ according to
eq.(\ref{Ri.p.t}) up to the factor $e^{-i\om t}$. This is exactly the result
expected from linear response theory\cite{kubo}.
\subsection*{B. Nonlinear modifications}
There are various contributions to the cubic response and therefore also
to the nonlinear modification of the response in a NHB field sequence.
These stem from the higher order terms in the expansion of the dipole
moment, eq.(\ref{mu.q}), and the anharmonic terms in the
potential, eq.(\ref{V.qi}).
In the lowest nonvanishing order, there are three relevant terms.
One term has its origin in the quartic anharmonicity, $\Theta_4$,
along with the first order term in the expansion of the dipole moment.
In the following, this term will be denoted as $\D P_i^{\rm(X)}(t,t_w,t_p)$.
In addition, there is one term, denoted by
$\D P_{i,(harm.)}^{\rm(X)}(t,t_w,t_p)$, which stems from the second order
term in eq.(\ref{mu.q}) and the harmonic contribution to $V(\{q\})$.
Finally, there exists one term which mixes the second order dipole moment term
with the one of the cubic anharmonicity, which we will call
$\D P_{i,(mix)}^{\rm(X)}(t,t_w,t_p)$.
The principal features of these terms are very similar with quantitative
differences only.

The different contributions are given by:
\Be\label{del.Pi}
\D P_i^{\rm(X)}(t,t_w,t_p)=&&\hspace{-0.6cm}
		-{3\over 2}{\rho\over m^3}\Theta_4E_ME_P^2
		\int\!d{\bf r}_0\int\!d{\bf r}_1
		\mu({\bf r}_0)_i'\mu({\bf r}_1)_i'e^{i{\bf k}_M{\bf r}_1}
		\times
		\nonumber\\
		&&\hspace{1.1cm}
		\times
		\int\!d{\bf r}_2\int\!d{\bf r}_3
		\mu({\bf r}_2)_i'\mu({\bf r}_3)_i'
			e^{i{\bf k}_P({\bf r}_2+{\bf r}_3)}
		\D R_i^{\rm(X)}(t,t_w,t_p)
\Ee
Here, $\D R_i^{\rm(X)}(t,t_w,t_p)$ denotes the intrinsic modification of
mode '$i$' and ${\bf k}_P$ and ${\bf k}_M$ are the wave vectors of the
pump-field and the measuring field, respectively.

From eq.(\ref{del.Pi}), the expression for
$\D P_{i,(harm.)}^{\rm(X)}(t,t_w,t_p)$ is obtained by dividing with
$(-{3\over 2}\Theta_4)$, replacing $\mu({\bf r}_k)_i'$, $k\!=\!1,2$ by
$\mu({\bf r}_k)_i''$ and using $\D R_{i,(harm.)}^{\rm(X)}(t,t_w,t_p)$ instead
of $\D R_i^{\rm(X)}(t,t_w,t_p)$.
In order to get the expression for $\D P_{i,(mix)}^{\rm(X)}(t,t_w,t_p)$ one has
to multiply $\D P_i^{\rm(X)}(t,t_w,t_p)$ in eq.(\ref{del.Pi}) by an overall
factor $(\Theta_3/\Theta_4)$. Furthermore, $\D R_i^{\rm(X)}(t,t_w,t_p)$ has to
be replaced by $\D R_{i,(mix)}^{\rm(X)}(t,t_w,t_p)$ and one of the three
$\mu({\bf r}_k)_i'$, $k\!\neq\!0$, by $\mu({\bf r}_k)_i''$.

The expressions for the modification $\D R_i^{\rm(X)}(t,t_w,t_p)$ (and also
$\D R_{i,(harm.)}^{\rm(X)}$, $\D R_{i,(mix)}^{\rm(X)}$) are somewhat more
complex than those for the linear response.
Here, we concentrate on $\D R_i^{\rm(X)}$ and only mention that the other
expressions are of a similar structure without giving them explicitly.

It turns out that for all three measuring procedures, X=S, P, and AC, the
modification can be written in the form:
\Be\label{del.Ri.X}
\D R_i^{\rm(X)}(t,t_w,t_p)=\left({\om_i^2\over\d_i^3}\right)
		  &&\hspace{-0.7cm}
		  \left\{
			e^{-2\l_{i,2}t_w}\hat{\chi}_{i,2}(\Om)^2
			h^{\rm X}(\l_{i,1},\l_{i,2};t)
		    - e^{-2\l_{i,1}t_w}\hat{\chi}_{i,1}(\Om)^2
			h^{\rm X}(\l_{i,2},\l_{i,1};t)
		  \right.\nonumber\\
                  &&\hspace{-0.6cm}
		    \left.
		     -e^{-(\l_{i,1}+\l_{i,2})t_w}
		      \hat{\chi}_{i,1}(\Om)\hat{\chi}_{i,2}(\Om)
		      g^{\rm X}(\l_{i,1},\l_{i,2};t)
		 \right\}
\Ee
In this expression, the functions $\hat{\chi}_{i,\a}(\Om)$ are defined by:
\be\label{chi.def}
\hat{\chi}_{i,\a}(\Om)
={\Om\over\l_{i,\a}^2+\Om^2}\left(1-e^{-\l_{i,\a}t_p}\right)
\quad;\quad \a=1,2 \quad;\quad t_p={2N\pi\over\Om}
\ee
with $N$ denoting the number of cycles of the pump-field.
The functions $h^{\rm X}(a,b;t)$ and $g^{\rm X}(a,b;t)$ are linear
combinations of exponential functions of time, given explicitly in the
appendix.
We mention that $\D R_i^{\rm(P/S)}(t,t_w,t_p)$ are real quantities despite the
fact that the $\l_{i,\a}$, $\a=1,2$, are complex in the underdamped case.

From the expression for the modifications $\D R_i^{\rm(X)}(t,t_w,t_p)$,
eq.(\ref{del.Ri.X}), it is evident that the absorbed energy plays an important
role, albeit the situation is more complex than in case of a completely
overdamped motion.
It can be shown from eq.(\ref{del.Ri.X}), using the expressions
given in the appendix, that the modification vanishes in the limit of small
as well as large times for $\g\!>\!0$ and thus is of a transient nature.
During the waiting time the created non-equilibrium population relaxes due to
the terms $e^{-\l_{i,\a}t_w}$. This is because there is no extra relaxation
mechanism in this model\cite{epl.nhb}.

Eq.(\ref{chi.def}) shows that
$\hat{\chi}_{i,\a}(\Om)\!\sim\!\Om$ for $\Om\!\to\!0$ and
$\hat{\chi}_{i,\a}(\Om)\!\sim\!\Om^{-2}$ for $\Om\!\to\!\infty$ and therefore
the modification vanishes in these cases also.
These considerations show that the nonlinear modification of the model
considered in ref.\cite{epl.nhb} and of the BO model are very similar,
particularly in the overdamped regime, which will be treated in more detail
in the next Section.

Because we mainly are interested in oscillatory motions, we consider the
signals in the frequency-domain as well. Here, the pulse-response
is of particular importance as this is the quantity that is directly related
to the complex dielectric constant in the linear regime.
Therefore, in addition to the time-domain signals we will consider the fourier
transforms of $R_i^{\rm(P)}(t)$ and $\D R_i^{\rm(P)}(t,t_w,t_p)$ with respect
to the measuring time $t$.
However, it is important to point out that the linear relation between the
pulse-response and the step-response,
$R_i^{\rm(S)}(t)=\int_0^t\!d\t R_i^{\rm(P)}(\t)$, {\it does not hold} for the
modifications $\D R_i^{\rm(S)}(t,t_w,t_p)$ and $\D R_i^{\rm(P)}(t,t_w,t_p)$
due to the nonlinear nature of these functions.
Similarly, there is no simple relation between the fourier transform of
$\D R_i^{\rm(P)}(t,t_w,t_p)$, denoted as $\D R_i^{\rm(P)}(\om,t_w,t_p)$ in the
following and $\D R_i^{\rm(AC)}(t,t_w,t_p)$. This can be inferred from the
definitions of the functions $h^{AC}$ and $g^{AC}$, eq.(\ref{h.g.ac.def}).
For long times the modification measured in the frequency domain,
$\D R_i^{\rm(AC)}(t,t_w,t_p)$, vanishes and there is no stationary
state evolving as it is the case in the linear response, cf.
eq.(\ref{Ri.ac.t}). This can be seen from eq.(\ref{h.g.ac.def})
as there is no term of the form $e^{-i\om t}$, but $\om$ occurs solely in
combinations with $\l_{i,\a}$. Of course, this behavior also is to be expected
intuitively because the modification is to be viewed as a transient effect.
It is for this reason that we consider $\D R_i^{\rm(P)}$ in the
frequency domain.
\subsection*{C. Overdamped limit: Ornstein-Uhlenbeck process}
As discussed above, the relation $dP^{\rm(S)}(t)/dt=P^{\rm(P)}(t)$
does not hold for the non-linear responses. This fact can be shown to hold
true for non-linear response functions in general.
Here, we will demonstrate it explicitly for the limiting case of a strongly
overdamped motion, the so-called Ornstein-Uhlenbeck (OU)-process\cite{vkamp81}.
All expressions for the OU-process can be obtained from the corresponding ones
for the BO-model from the lowest-order term of an expansion in $1/\g_i$.
This expansion can either be performed on the Fokker-Planck equation itself
or on any of the resulting expressions for the linear or nonlinear response.
For the OU-process, the effective relaxation rate is given by
$\L_i=\om_i^2/\g_i$\cite{mukamel}.
Additionally, the expressions for $R_i^{\rm(S)}$ and $\D R_i^{\rm(S)}$ obtained
that way allow a direct comparison with those obtained for the model of
stochastic dipole reorientations in ref.\cite{epl.nhb}.
The linear response given in eq.(\ref{Ri.p.t}) simplifies to
\[
R_i^{\rm(P)}(t)_{\rm(OU)}={\L_i\over \om_i^2}e^{-\L_it}
\quad\mbox{and}\quad
R_i^{\rm(S)}(t)_{\rm(OU)}={1\over \om_i^2}\left(1-e^{-\L_it}\right)
\]
For the modifications, however, one finds from eq.(\ref{del.Ri.X}):
\Be\label{del.Ri.OU}
\D R_i^{\rm(P)}(t,t_w,t_p)_{\rm(OU)}=&&\hspace{-0.6cm}
			    {\L_i\over\om_i^6}A_i(\Om)e^{-2\L_it_w}
			      \left(1-e^{-2\L_it}\right)e^{-\L_it}
			    \nonumber\\
\D R_i^{\rm(S)}(t,t_w,t_p)_{\rm(OU)}=&&\hspace{-0.6cm}
			    {1\over\om_i^6}A_i(\Om)e^{-2\L_it_w}
			      \left(1-e^{-\L_it}\right)^2e^{-\L_it}
\Ee
Here, the pump-frequency dependent amplitude is given by
$A_i(\Om)=\left[\e_i''(\Om)\right]^2\left(1-e^{-\L_it_p}\right)^2$ with
$\e_i(\om)=\L_i/(\L_i-i\om)$.
It is evident from these expressions that
$\D R_i^{\rm(S)}(t,t_w,t_p)\neq\int_0^t\!d\t\D R_i^{\rm(P)}(\t,t_w,t_p)$.
Furthermore, the expression for $\D R_i^{\rm(S)}(t,t_w,t_p)_{\rm(OU)}$ is
similar to the one found in ref.\cite{epl.nhb} for stochastic dipole
reorientations apart from some minor differences, which will be discussed in
detail elsewhere\cite{uli.ou}.

If we consider the fourier transform of $R_i^{\rm(P)}(t)_{\rm(OU)}$,
$R_i^{\rm(P)}(\om)_{\rm(OU)}=\om_i^{-2}\e_i(\om)$ and compare it to the signal
measured in the frequency domain via application of a field $E_Me^{-i\om t}$,
\[
R_i^{\rm(AC)}(t)_{\rm(OU)}=
\om_i^{-2}\e_i(\om)\left(e^{-i\om t}-e^{-\L_it}\right),
\]
for $t\!\gg\!\L_i^{-1}$ one finds the standard result of linear response
theory, namely
$R_i^{\rm(AC)}(t)_{\rm(OU)}=e^{-i\om t}R_i^{\rm(P)}(\om)_{\rm(OU)}$.
However, this does not hold for the modification
$\D R_i^{\rm(AC)}(t,t_w,t_p)_{\rm(OU)}$.
According to eq.(\ref{del.Ri.X}), we have:
\[
\D R_i^{\rm(AC)}(t,t_w,t_p)''_{\rm(OU)}=
{1\over \om_i^6}A_i(\Om)e^{-2\L_it_w}\epsilon_i^{''}(\omega)
\left[2\L_ie^{-2\L_i t}{\sin{(\om t)}\over\om}+(e^{-3\L_i t}-e^{-\L_i t})\right]
\]
which has to be compared to the imaginary part of
\be\label{del.Ri.OU.om}
\D R_i^{\rm(P)}(\om,t_w,t_p)_{\rm(OU)}=
{1\over 3\om_i^6}A_i(\Om)e^{-2\L_it_w}\left[3\e_i(\om)-\e_i(\om/3)\right]
\ee
Whereas the dependence on the burn frequency $\Om$ is identical, this does
obviously not hold with respect to the frequency $\om$.
The modification $\D R_i^{\rm(AC)}(t,t_w,t_p)''_{\rm(OU)}$ clearly exhibits
a transient behavior with respect to the time $t$ as it vanishes in the limit
of short and long times.
There is no stationary state which would allow to set a time-window for
observation, in marked contrast to the linear response.
These considerations show that it is most meaningful to consider
$\D R_i^{\rm(P)}(\om,t_w,t_p)$, i.e. the fourier transform of
$\D R_i^{\rm(P)}(t,t_w,t_p)$, in the frequency domain and {\it not} the signal
measured via application of an ac-field.
\subsection*{D. Light to vibration coupling}
As already noted above, in amorphous systems spatial correlations of the
dipole moments are expected to exist which extent over some finite volume,
depending on the coherence length of the modes $q_i$.
Therefore, the position-dependence of the $\mu({\bf r})_i'$ and the
$\mu({\bf r})_i''$ plays an important role.
This fact complicates the relation between the intrinsic response functions
and the respective polarizations, eqns.(\ref{Pi.lin}) and (\ref{del.Pi}).
Thus, the situation is very similar to the case of Raman scattering in
glasses\cite{SG70}.

In the expression for the linear response, eq.(\ref{Pi.lin}), we have to
consider the correlation function
$\int\!d{\bf r}_0\int\!d{\bf r}_1\mu({\bf r}_0)'_i\mu({\bf r}_1)_i'$,
where we have used $e^{i{\bf k}_M{\bf r_1}}\!\simeq\!1$, i.e. the
${\bf k}\!\to\!0$ limit of
$\lg\mu(-{\bf k})_i'\mu({\bf k})_i'\rg$.
In order to proceed in the calculation of this function, we make the same
assumptions as they are typically used in calculations of the Raman-scattering
intensity from amorphous systems\cite{SG70, MB74}.
We assume that the coherence length of the mode $q_i$ is much smaller than
the wavelength of the light.
Additionally, we disregard the dependence of all quantities on the polarization
of the mode $q_i$ and of the light.
In an amorphous system $\lg\mu(-{\bf k})_i'\mu({\bf k})_i'\rg$ is expected to
have a broad flat maximum in the vicinity of ${\bf k}=0$, in vast contrast to
the situation in crystals.
This is the reason for the breaking of the momentum transfer selection rules
in the case of Raman-scattering\cite{SG70}.
A comparison of our expression to the ones obtained for the
Raman-scattering intensity shows that $\lg\mu(-{\bf k})_i'\mu({\bf k})_i'\rg$
is proportional to the so-called light to vibration coupling $C(\om_i)$.
Indeed, a very similar behavior of the Raman- and Infrared-coupling constants
has been observed in a silica glass\cite{Ahmad93}.
We thus assume $\lg\mu(-{\bf k})_i'\mu({\bf k})_i'\rg\!\sim\!C(\om_i)$ apart
from an overall prefactor $|\mu_i'|^2$, which replaces $|\a_i'|^2$ occuring in
the Raman case, i.e.
\be\label{mu.cf.lin}
\lg\mu(-{\bf k})_i'\mu({\bf k})_i'\rg\!\simeq\!|\mu_i'|^2C(\om_i)
\ee
In the expression for the modification $\D P_i^{(X)}$, eq.(\ref{del.Pi}),
the fourier transform of the four-point correlation function
$\lg\mu({\bf r}_0)_i'\mu({\bf r}_1)_i'\mu({\bf r}_2)_i'\mu({\bf r}_3)_i'\rg$
occurs.
In general such quantities are extremely difficult to calculate.
Therefore, in order to be able to proceed we assume that the probability
generating functional for the correlations can be represented by a Gaussian.
Note that this is exact for harmonic vibrations and otherwise represents
a mean-field approximation for the spatial correlations of the modes $q_i$.
With this approximation the four-point correlations factor into products of
two-point correlations yielding
\[
\int\!d{\bf r}_0\int\!d{\bf r}_1\int\!d{\bf r}_2\int\!d{\bf r}_3
\mu({\bf r}_0)_i'\mu({\bf r}_1)_i'\mu({\bf r}_2)_i'\mu({\bf r}_3)_i'
e^{i[{\bf k}_M{\bf r}_1 + {\bf k}_P({\bf r}_2+{\bf r}_3)]}
\!\simeq\!3|\mu_i'|^4C(\om_i)^2
\]
in the expression for $\D P_i^{(X)}$, eq.(\ref{del.Pi}).

For $\D P_{i,(harm.)}^{(X)}$ another four-point correlation function,
$\lg\mu({\bf r}_0)_i'\mu({\bf r}_1)_i''\mu({\bf r}_2)_i''\mu({\bf r}_3)_i'\rg$
occurs, cf. the discussion in the context of eq.(\ref{del.Pi}).
In the mean-field approximation one thus has to deal not only with
$\lg\mu(-{\bf k})_i'\mu({\bf k})_i'\rg$, but also with the unkown correlation
functions $\lg\mu(-{\bf k})_i'\mu({\bf k})_i''\rg$ and
$\lg\mu(-{\bf k})_i''\mu({\bf k})_i''\rg$.
For the function $\lg\mu(-{\bf k})_i'\mu({\bf k})_i''\rg$ it is reasonable
to assume that it is vanishingly small in isotropic systems because of the
odd number of derivatives. We thus assume
\be\label{mu.cf.mix}
\lg\mu(-{\bf k})_i'\mu({\bf k})_i''\rg\simeq 0
\ee
For the function $\lg\mu(-{\bf k})_i''\mu({\bf k})_i''\rg$ one expects a
similar behavior as for $\lg\mu(-{\bf k})_i'\mu({\bf k})_i'\rg$. In general,
however, the frequency-dependences may be different.
Therefore, we write
\be\label{mu.cf.sq}
\lg\mu(-{\bf k})_i''\mu({\bf k})_i''\rg\!\simeq\!|\mu_i''|^2C'(\om_i)
\ee
and this way find
\[
\lg\mu({\bf r}_0)_i'\mu({\bf r}_1)_i''\mu({\bf r}_2)_i''\mu({\bf r}_3)_i'\rg
\!\simeq\!|\mu_i'|^2|\mu_i''|^2C(\om_i)C'(\om_i)
\]
Only if one assumes in an ad hoc manner that the $\mu({\bf r})_i''$ couple to
the same elasto-optical constants as the $\mu({\bf r})_i'$ one has
$C'(\om_i)\!=\!C(\om_i)$ in the expression for $\D P_{i,(harm.)}^{(X)}$.
Due to the lack of a theoretical argument in favor of such an approximation
we mainly focus on $\D P_i^{(X)}$ throughout the present paper.

The last term occuring in eq.(\ref{del.Pi}) is the 'cross-term'
$\D P_{i,(mix)}^{(X)}$.
The relevant four-point correlation function in this context is
$\lg\mu({\bf r}_0)_i'\mu({\bf r}_1)_i'\mu({\bf r}_2)_i''\mu({\bf r}_3)_i'\rg$.
However, in every term of the mean-field approximation the two-point
correlation function $\lg\mu(-{\bf k})_i'\mu({\bf k})_i''\rg$ occurs.
Therefore, due to eq.(\ref{mu.cf.mix}) we approximately have
\[
\D P_{i,(mix)}^{(X)}(t,t_w,t_p)\simeq 0
\]

Various models for the low-frequency excitations in glasses have been used
to calculate $C(\om_i)$. All of them yield power laws, $C(\om_i)\sim\om_i^n$,
with exponents $n$ ranging from $n\!=\!0$ in the soft potential
model\cite{GKK89} to $n\!=\!2$ in the harmonic model\cite{MB74}.
Furthermore, when $C(\om_i)$ is analyzed via a comparison between experimental
Raman- and neutron-scattering intensities from the same sample\cite{MNPSZ90}
often a $C(\om_i)\sim\om_i$ dependence is found.
Also a nonvanishing limiting value $C(\om_i\to 0)\neq 0$ has been
reported\cite{Fetal99}.
It also should be mentioned that even an explicit ${\bf k}$-dependence of the
low-frequency Raman-scattering intensity has been observed in a silica
glass\cite{SWNRD99}.
Thus, given the uncertainty regarding the light to vibration coupling, in the
present paper we will use $C(\om_i)\sim\om_i^n$ with $n=0,1,2$ in
order to discuss the possible behavior.

In actual calculations, we furthermore disregard the dependence of
$\mu({\bf r})_i'$ on the mode-index $i$ and write for the coupling constant
\be\label{C.om}
\lg\mu(-{\bf k})_i'\mu({\bf k})_i'\rg
=|\mu'|^2C(\om_i)\quad\mbox{with}\quad C(\om_i)\propto\om_i^n
\ee
It should be noted that if the system under consideration consists of
independent particles, spatial correlations of $\mu$ vanish and the relation
between the polarization and the response functions become trivial.
In this case we simply have $C(\om_i)\!=\!C'(\om_i)\!=\!1$ and only the
corresponding prefactors $|\mu_i'|^2$ and $|\mu_i''|^2$ occur in the
expressions for the polarization.
\subsection*{E. Polarization}
In addition to the approximations discussed above we assume that the parameters
of the BO-model, $\g_i$ and $\om_i$, are distributed according to some
distribution functions $g(\om_i)$ and $p(\g_i)$, which we choose to be
independent.
For the density of states (DOS) of the low-frequency vibrations we will use
$g(\om_i)\sim\om_i^m$, $m=2,4$ as they follow for the Debye model and the
soft potential model, respectively.
For computational convenience we additionally introduce a high-frequency
cut-off $\om_c$:
\be\label{dos}
g(\om_i)=\left({m+1\over \om_c^{m+1}}\right)\om^m\theta(\om_c-\om_i)
\ee
With these approximations and definitions we have from eq.(\ref{Pi.lin}):
\be\label{P.ges}
P^{\rm(X)}(t)\simeq{\rho\over m} E_M|\mu'|^2
		     \int\!d\om_i\!\int\!d\g_ig(\om_i)p(\g_i)
					   C(\om_i)R_i^{\rm(X)}(t)
\ee
and an analogous expression for the modification, eq.(\ref{del.Pi}):
\be\label{del.P.ges}
\D P^{\rm(X)}(t,t_w,t_p)\simeq
			 -{3\over 2}{\rho\over m^3}\Theta_4E_ME_P^2|\mu'|^4
				  \int\!d\om_i\!\int\!d\g_i g(\om_i)p(\g_i)
				  \left[3C(\om_i)^2\right]
				  \D R_i^{\rm(X)}(t,t_w,t_p)
\ee
As already noted in connection with eq.(\ref{del.Pi}) and in the previous
section, the expression for $\D P_{(harm.)}^{\rm(X)}$ is obtained from the
one for $\D P^{\rm(X)}$ by replacing
$[-{9\over 2}\Theta_4|\mu'|^4C(\om_i)^2]$ by
$[|\mu'|^2|\mu''|^2C(\om_i)C'(\om_i)]$ and using $\D R_{i,(harm.)}^{\rm(X)}$
instead of $\D R_i^{\rm(X)}$.
These expressions will be used to discuss the response of a collection of
Brownian oscillators to the NHB pulse sequence.

At this point it should be pointed out that in case of $\Theta_4\!>\!0$ the
sign of the two functions $\D P^{\rm(X)}$ and $\D P_{(harm.)}^{\rm(X)}$
will be the opposite.
This is an example of the fact that one cannot predict the sign of a
third-order response in general.
\section*{III. Results and Discussion}
In this section we will discuss the results of model calculations and show
that indeed the NHB field sequence is capable to detect dynamic
heterogeneities provided they exist. In the discussion of the nonlinear
modification we will mainly focus on $\D P^{\rm(X)}$ but also show results
for $\D P_{(harm.)}^{\rm(X)}$. In addition we will give an order of magnitude
estimate of the expected effects, as this is of utmost importance for possible
experimental realizations.
\subsection*{A. Dielectric loss}
We start with a brief discussion of the dielectric loss, $\e''(\om)$.
From eq.(\ref{P.ges}) and the fourier transform of eq.(\ref{Ri.p.t}) one
obtains $\e''(\om)=(\e_0E_M)^{-1}{\cal F}[P^{\rm(P)}(t)]$, where $\e_0$ is the
permittivity of free space.
For simplicity we restrict ourselves to a single damping constant,
i.e. we use $p(\g_i)=\d(\g_i-\g)$ in eq.(\ref{P.ges}). We then explicitly
have:
\be\label{eps.om}
\e''(\om)={\rho\over\e_0 m}|\mu'|^2\int\!d\om_ig(\om_i)C(\om_i)
	  {\g\om\over (\om_i^2-\om^2)^2+(\g\om)^2}
\ee
The limiting behavior of $\e''(\om)$ can easily be extracted from this
expression.
For vanishing $\g$ eq.(\ref{eps.om}) reduces to
$\e''(\om)=\rho(\e_0m)^{-1}(\pi/2)|\mu'|^2C(\om)g(\om)/\om$  which is
directly proportional to the Raman intensity scaled to the Bose factor.
Using eq.(\ref{C.om}) for $C(\om_i)$ and eq.(\ref{dos}) for $g(\om_i)$ this
implies $\e''(\om)\propto\om^{m+n-1}$.
For finite $\g$ we have $\g>\om$ at low frequencies and therefore
$\e''(\om)\sim\om$.
For frequencies $\om>\g$ this behavior changes into that of undamped
oscillations.
Since the DOS has a high frequency cut-off $\om_c$ we find for
$\om>\om_c$ $\e''(\om)\sim\om^{-1}$ in the overdamped case and
$\e''(\om)\sim\om^{-3}$ otherwise.
This behavior is exemplified in Fig.2, where we plotted $\e''(\om)$ versus
$\om/\om_c$ for $m=2$, $n=1$ and various values of the damping constant $\g$.
Also included in Fig.2 as the thin dotted line is the dielectric loss for a
Debye relaxation, $\e_D''(\om)=\g\om/(\om^2+\g^2)$, which behaves as
$\e_D''(\om)\simeq\g^{-1}\om$ and $\e_D''(\om)\simeq\g\om^{-1}$ for
small and large $\om$, respectively. This shows that in the overdamped case we
have an apparent distribution of relaxation times,
giving rise to a sublinear increase of $\e''(\om)$ for small $\om$.
Of course, the cusp in $\e''(\om)$ for small $\g$ has its origin solely in the
assumed high-frequency cut-off of $g(\om)$.
\subsection*{B. Nonlinear modifications}
Before we discuss the behavior of the modification
$\D P^{\rm(P)}(\om,t_w,t_p)$ it is instructive to consider the response
associated with a single mode $q_i$.
In Fig.3 we show the real and imaginary parts of
$\D R^{\rm(P)}_i(\om,t_w,t_p)$, the fourier transform of
$\D R^{\rm(P)}_i(t,t_w,t_p)$ given in eq.(\ref{del.Ri.X}), for a burn
frequency $\Om=1.0$.
The full lines correspond to an underdamped oscillator whereas the dashed line
represents the overdamped case.
Note that ${\hat \chi}_{i,\a}(\Om)$ in this situation only determines the
overall amplitude of the modification. Only when a distribution of modes
is considered the frequency selectivity of ${\hat \chi}_{i,\a}(\Om)$
becomes important.
In the present situation of a single mode this only means that the overall
amplitude of $\D R^{\rm(P)}_i(t,t_w,t_p)$ is changed if the burn frequency
$\Om$ is varied.
In the underdamped case a clear resonance occuring at $\om=\om_i$ is visible
in both, $\D R^{\rm(P)}_i(\om,t_w,t_p)''$ and
$\D R^{\rm(P)}_i(\om,t_w,t_p)'$.
Additionally, a weak resonant behavior is found at $\om=3\om_i$.
This can be understood from the definition of $h^P$ and $g^P$, cf.
eq.(\ref{h.g.p.def}), and stems from terms of the form $1/[3\l_{i,\a}-i\om]$.
On the other hand, in the overdamped case the maximum modification is
found at $\om\sim\L_i$ (more precisely $\om\simeq 0.86\L_i$).
In the limit of the OU-process, the result for $\D R_i^{\rm(P)}$ has been given
above in eq.(\ref{del.Ri.OU.om}).
The factor $\om_i^6$ in the denominator explains the smallness of
$\D R_i^{\rm(P)}(\om,t_w,t_p)_{\rm(OU)}$, which in Fig.3 was multiplied by a
factor $10^8$.
In this case of overdamped dynamics the imaginary part appears to be better
suited to find the maximum modification.
Thus, in the following we concentrate on $\D R^{\rm(P)}_i(\om,t_w,t_p)''$
although in some situations the real part may also yield helpful information.

The behavior of $\D R_i^{\rm(P)}(\om,t_w,t_p)$ as a function of $\om$ also
determines the modification of the polarization, $\D P^{\rm(P)}(\om,t_w,t_p)$
according to eq.(\ref{del.P.ges}).
In the upper panel of Fig.4a we plotted
$\D P^{\rm(P)}(\om)''\equiv\D P^{\rm(P)}(\om,t_w,t_p)''$ versus frequency for
various pump-frequencies $\Om$ and one cycle of the pump-field.
As in Fig.2, we chose $m\!=\!2$ and $n\!=\!1$, i.e.
$g(\om_i)\!\propto\!\om_i^2$, $C(\om_i)\!=\!\om_i$.
Additionally, we consider a single damping constant, thus writing
$p(\g_i)\!=\!\d(\g_i-\g)$ with $\g\!=\!10^{-2}\om_c$ in eq.(\ref{del.P.ges}).
We find that for all $\Om$ the minimum of the modification is observed at
frequencies $\om_{\rm min}\!\sim\!\Om$.
In the lower panel $\D P_{(harm.)}^{\rm(P)}(\om)''$ is shown.
In this calculation we assumed $C'(\om_i)\!=\!C(\om_i)$ for simplicity and
used the same parameters as in the upper panel.
It is evident that the main differences are the reversed sign of the
modification and that $\om_{\rm max}$ is slightly larger for all pump
frequencies.
Additionally, in the underdamped regime a slight minimum is observable at
$\om\!=\!\g$, which is absent in $\D P^{\rm(P)}(\om)''$.
However, the most important conclusion that can be drawn from Fig.4a is that
the modification of the response clearly exhibits a pronounced
$\Om$-selectivity.
This means, that NHB is capable to 'detect' dynamic heterogeneities also in
the regime of oscillatory motions.

In addition to the $\Om$-selectivity we can discriminate between underdamped
and overdamped modes due to the functional form of $\D P^{\rm(P)}(\om)''$.
In the overdamped regime the modification only shows a minimum, cf.
eq.(\ref{del.Ri.OU.om}) (a maximum in case of
$\D P_{(harm.)}^{\rm(P)}(\om)''$).
In the underdamped case an additional dispersive behavior is observed.
This means that also in the macroscopic polarization the main features already
observed for the response of a single mode persist.
In Fig.4a, the dashed lines correspond to overdamped modes because here
$\Om\ll\g$ and thus mainly the response of those modes with $\L_i\!\sim\!\Om$
is modified by the pump.
For higher burn frequencies, however, mainly the response of underdamped modes
is modified, cf. the full lines in Fig.4a.
Here, the maximum modification is around $\om\!\sim\!\om_i$ and therefore
$\om_i\!\simeq\!\Om$.
Of course, for other choices of the damping constant $\g$ the cross-over
from underdamped to overdamped modes takes place at a different frequency.

If we compare the form of the modifications in the underdamped regime,
the full lines in the upper panel of Fig.4a, to those for a single underdamped
mode shown in Fig.3, it is evident immediately that the range of non-vanishing
modification extents over a considerably larger $\om$-range.
This has its origin mainly in the functional form of $\hat{\chi}_{i,\a}(\Om)$,
cf. eq.(\ref{chi.def}), in particular the factor $(1-e^{-\l_{i,\a}t_p})$, which
shows oscillatory behavior in the underdamped case.
At low frequencies some more strongly damped modes contribute to the signal,
rendering the modification broader as compared to Fig.3.
This effect can be suppressed by increasing the number of cycles $N$
of the sinusoidal pump-field.
The condition $e^{-\l_{i,\a}t_p}\!\ll\! 1$ is fulfilled for
$N\!\gg\!\Om/(\pi\g_i)$ and thus for a large number of cycles the behavior
of a single oscillator is approached.
Therefore, in the regime of underdamped modes the number of cycles $N$ may
turn out to be a useful experimental parameter.
This is in contrast to the situation of overdamped modes where increasing $N$
mainly gives rise to a small increase in the overall amplitude due to the
factor $(1-e^{-\L_it_p})$.

Next, we consider the frequency at the minimum modification, $\om_{\rm min}$.
This quantity is plotted versus $\Om$ in Fig.4b for two pairs of $(m,n)$.
Remember that these parameters determine the effective distribution of
$\om_i$ contributing to the signal because according to eq.(\ref{del.P.ges})
we have $g(\om_i)C(\om_i)^2\!\propto\!\om_i^{m+2n}$.
Due to this behavior, the equivalences $(2,1)\!\doteq\!(4,0)$ and
$(2,2)\!\doteq\!(4,1)$ hold for the pairs $(m,n)$ shown.
As in Fig.4a, the upper panel shows $\om_{\rm min}$ extracted from
$\D P^{\rm(P)}(\om)''$ and in the lower panel $\om_{\rm max}$ obtained from
$\D P_{(harm.)}^{\rm(P)}(\om)''$ is plotted. In the latter case, we again
assumed $C'(\om_i)\!=\!C(\om_i)$.
It is evident from Fig.4b that $\om_{\rm min}\!\propto\!\Om$.
Only in the regime of overdamped motion different pairs ($m$,$n$) give rise
to some minor differences in the values of $\om_{\rm min}$ ($\om_{\rm max}$).
Of course, the linear relation $\om_{\rm min}\!\propto\!\Om$ has its origin in
the form of the effective distribution of $\om_i$ in the oscillatory regime
and the corresponding distribution of $\L_i\!=\!\om_i^2/\g$ in the overdamped
regime.
In the underdamped case, one does not only have a minimum in
$\D P^{\rm(P)}(\om)''$ but also a maximum in addition, cf. Fig.4a.
However, as discussed above, with increasing number of cycles, the differences
between the minimum- and maximum-positions diminishes.

Sofar, we have considered the case of a fixed damping constant, which of
course should be interpreted as an average value.
It is evident from the discussion of Fig.4 that the value of $\g$ can hardly
be extracted from $\D P^{\rm(P)}(\om,t_w,t_p)''$ and this does not change if
a distribution of damping constants is assumed in the calculations.
On the other hand, it is known from previous studies that
$\D P^{\rm(S)}(t,t_w,t_p)$ is sensitive to relaxation rates and we thus
expect the same for $\D P^{\rm(P)}(t,t_w,t_p)$.
In order to demonstrate that in the oscillatory regime measuring
$\D P^{\rm(P)}(t,t_w,t_p)$ may provide information about the damping constants,
in the upper panel of Fig.5 we plot this quantity versus measuring time for the
same parameters as used in Fig.4.
Here, we used $\Om/\om_c=0.1$ and various numbers of cycles of the pump-field,
$N$.
The first thing to notice is that for $N\!=\!1$ hardly any oscillations are
visible in the time-domain, whereas this behavior changes with increasing $N$.
This fact has the same origin as the decreasing broadening of the modifications
for larger $N$, as discussed in context with Fig.4a.
For a single cycle of the pump-field $\D P^{\rm(P)}(\om,t_w,t_p)''$ shows a
very broad resonance and accordingly the dot-dashed line shows almost no
oscillations.
For $N\!=\!10$ $\D P^{\rm(P)}(t,t_w,t_p)$ shows the expected oscillatory
behavior.
More important for the present discussion, however, is the fact that for
a large number of cycles it is evident that the signal has its maximum
envelope around $t\!\sim\!\g^{-1}$.
This is demonstrated by the dotted lines, which are of the form
$t\!\times\!e^{-\g t}$.
This functional form of the envelope can be derived analytically from the
expression for the pulse-response of a single mode, eq.(\ref{del.Ri.X}).

The lower panel of Fig.5 shows $\D P_{(harm.)}^{\rm(P)}(t,t_w,t_p)$ for the
same parameters and the additional assumption $C'(\om_i)\!=\!C(\om_i)$.
In this case, however, the situation is quite different.
Here the main effects are observed at $\g t\!=\!\Om$ and the oscillatory
behavior around $\g t\!\simeq\!1$ is strongly suppressed.
On the other hand, the maxima of the signal are by about a factor of
ten smaller than in case of $\D P^{\rm(P)}(t,t_w,t_p)$. (The envelope
function (dotted line) is the same as in the upper panel but divided by
a factor of $30$.)
Thus, we conclude that also in the oscillatory regime NHB experiments may
be useful in order to get information about the damping, although this might
require to perform experiments with various numbers of pump-field cycles.
In particular, this should be feasible if the main contribution to the signal
stems from $\D P^{\rm(P)}(t,t_w,t_p)$, i.e. for very small $|\mu''|^2$.
\subsection*{C. Estimated magnitude of nonlinear effects}
For an experimental realization of the NHB experiment in the THz regime it
is of course desirable to have at least an order of magnitude estimate of
the expected nonlinear effects.
For this reason we consider the ratio $\D P/P$, where $P$ and $\D P$ are
shorthand notations for the linear response and the nonlinear modification.
Experimentally, the sum $P^*=P+\D P$ will be observed and $\D P$ has to be
extracted by a proper phase cycling procedure, cf. ref.\cite{nhb.sci}.

We start our consideration with $\D P(\om)''$ for a single overdamped mode,
given in eq.(\ref{del.Ri.OU.om}). For our rough estimate we consider the
frequency of the maximum modification, i.e. $\Om\!=\!\L_i\!=\!\om$, which
allows us to neglect $\e_i''(\L_i/3)$ as compared to $\e_i''(\L_i)$ and to
use $\exp (-\L_i t_p)\!\simeq\!0$.
Furthermore we set the waiting time to zero.
This way we find:
\[
\frac{\D P_i}{P_i}\simeq\frac{3}{8}\frac{E_p^2|\mu_i'|^2}{m^2\om_i^4}\Theta_4
\]
In order to proceed we replace the derivative of the dipole moment by
$\mu_i'\simeq(\d\mu/\d q_i)$ and assume that $\left(\d q_i\right)^2$ is either
related to the thermal expectation value
$\lg q_i^2\rg\!=\!(k_B T)/(m\om_i^2)$ or the average displacement
${\bar q}_i^2$.
This allows us to write
\be\label{dP.P.OU.estimate}
\frac{\D P}{P}\simeq\frac{3}{8}\left(\frac{\d\mu\cdot E_P}{k_BT}\right)^2
\bar{q}_i^2\Theta_4
\ee
In order to proceed, we need an estimate of the quartic anharmonicity
$\Theta_4$. For this purpose we utilize the detailed comparison of the
predictions of the soft-potential model (where a potential of a similar form
as in eq.(\ref{V.qi}) is used) with experimental results on a variety of
different glassforming liquids\cite{Parshin94}.
From this comparison we find an average value of
$\Theta_4\!\simeq\!2\cdot 10^{22}m^{-2}$.
Assuming a value of $|\mu_i'|$ on the order of $1$Db/$1$\AA\cite{HB89}, the
value of the actual displacement and the one to be chosen for $\d\mu$ are not
independent of each other.
If $\bar{q}_i\!=\!x\times 1$\AA\  we should use $\d\mu\!\simeq\!x\times 1$Db.
Note that this gives an estimate for
$\bar{q}_i^2\Theta_4\!\simeq\!2x^2\cdot 10^2$, which should be smaller than
unity in order for the perturbation expansion used in Sect.II to be meaningful.
Assuming a pump-field amplitude $E_P\!=\!z\times 10^4$kV/cm one then finds,
using ($10^4$kV/cm$\times$1Db)/($k_B\!\times\!100$K)$\simeq 2.5$,
that $\D P/P\!\simeq 5\cdot10^2\cdot x^4z^2$. Thus, for a reasonable value of
$x\!\simeq\!0.05$ one has $\D P/P\!\simeq 3\cdot 10^{-3}z^2$, meaning that
$\D P/P$ is in the percent range for $z\!\sim\!2$ or
$E_P\!\simeq\!2\cdot10^4$kV/cm.

In this context, it should be mentioned, that the above estimate of course
should not be used if one is concerned with the primary relaxation of high
temperature liquids. This is because in that case the dielectric relaxation is
mainly determined by the mean square fluctuations, $\lg\D\mu^2\rg$, of the
static molecular dipole moment $\mu$ due to the molecular tumbling motion.
In our formulation this means that we loosely should identify
$\left(\d\mu\bar{q}_i\right)^2\Theta_4$ with $\lg\D\mu^2\rg$,
cf. ref.\cite{epl.nhb}.
Assuming isotropic rotational jumps as a simple but realistic model for the
stochastic reorientational motion in liquids\cite{DBHH98} allows us to
identify $\bar{q}_i$ with a 'rotational jump length',
$\bar{q}_i\!\simeq\!R_H\sin{(\phi_{rot})}$. Here, the hydrodynamic radius
$R_H$ is on the order of 1\AA\cite{CS97}, yielding
$\bar{q}_i^2\Theta_4\!\simeq\!6$, if the mean jump angle $\phi_{rot}$ is on
the order of ten degrees.
Again we write $E_P\!=\!z\times 10^4$kV/cm in eq.(\ref{dP.P.OU.estimate}),
but now we use $T\!=\!300$K. For a dipole moment of $1$Db this gives
$\left(\D P/P\right)_{(\rm{rot.},300 \rm K)}\!\simeq\!4z^2$.
Therefore, a pump-field amplitude of $E_P\!\simeq500$kV/cm is sufficient
to get an effect in the percent range. This value is in harmony with the
amplitudes used in the NHB experiments on supercooled liquids\cite{nhb.sci}.

Next, we consider the more relevant case of underdamped oscillatory motion.
In order to give an estimate of the ratio $\D P(\om)''/P(\om)''$ in this
regime we now consider a distribution of eigenfrequencies $\om_i$ similar to
the discussion in context with Fig. 4a (upper panel) and
$C(\om_i)\!\propto\!\om_i$.
The expression for the dielectric loss, eq.(\ref{eps.om}), allows us to
write
\be\label{Dp.P.oszi}
\frac{\D P(\om)''}{P(\om)''}=
\frac{9}{2}\e_0 \e''(\om)\rho_m^{-1}E_p^2
\Theta_4 \frac{\int\!d\om_i g(\om_i)\om_i^2\D R_i^{(P)}(\om)''}
	      {\left[\int\!d\om_ig(\om_i)\om_iR_i^{(P)}(\om)''\right]^2}
\ee
In this expression, $\rho_m\!=\!m\rho\!\simeq\!10^3 kg/m^3$ is the samples
mass density.
In order to obtain a peak in the THz-range we now use a DOS of the form
$g(\om_i)\!\propto\!\om_i^2\exp(-\t\om_i)$, with
$\t\!=\!4/3\cdot 10^{-12}$ Hz$^{-1}$, yielding a peak at $\om_i\!=\!1.5$THz.
Consequently, we set $\Om\!=\!1.5$THz.
Additionally, we fix the damping constant to value of $\g\!=\!0.1$THz.
We point out, that a variation of $\g$ has only a minor effect on the resulting
amplitudes as long as the motion is in the strongly underdamped regime.
For the dielectric loss in the THz-range we approximately have
$\e''(\mbox{THz})\!\simeq\!1$, as found for glycerol\cite{Lunki96}.
Evaluating the integrals in eq.(\ref{Dp.P.oszi}) at
$\om_{\rm min}\!\simeq\!1$THz (cf. Fig.4) allows us to calculate
$\D P_{\rm max}$. For $\Theta_4$ we use the same value as before,
$\Theta_4\!\simeq\!10^{22} m^{-2}$ and find
\[
\left|\frac{\D P_{\rm max}}{P} (1\rm{THz})\right|\simeq 3\cdot10^{-15}
\frac{E_p^2}{\left(\rm{V/m}\right)^2}
\]
This way we obtain $\D P_{\rm max}/P\!\simeq\!0.3$ for a pump-field amplitude
of $E_P=10^2$kV/cm.
These considerations allow us to conclude that - with the assumptions used -
the ratio $\D P/P$ should be larger in the oscillatory regime than
in the overdamped regime.
Of course, from an intuitive point of view this does not come as a surprise
because some resonant behavior even is expected in this case.

Sofar, we have considered the nonlinear modification $\D P(\om)''$.
We close this section with a brief discussion of the relative magnitudes
of $\D P$ and $\D P_{(harm.)}$, the nonlinear modification stemming from the
second order term in the expansion of the dipole moment, eq.(\ref{mu.q}).
As is evident from Fig.4a, the magnitudes of the normalized effects are quite
similar.
Thus, we have to compare the prefactors $3\Theta_4|\mu'|^2$ and $|\mu''|^2$.
For this purpose we proceed in exactly the same way as we did in the
discussion of the overdamped motion. We approximate
$|\mu'|^2$ by $(\d\mu)^2/\lg q^2\rg$ and $|\mu''|^2$ by
$(\d[\d\mu])^2/(3\lg q^2\rg\bar{q}^2)$.
This way we find for the ratio
\[
{3\Theta_4|\mu'|^2\over|\mu''|^2}\simeq
9\bar{q}^2\Theta_4\left({\d\mu\over\d[\d\mu]}\right)^2
\sim 2 x^2\cdot 10^3\left({\d\mu\over\d[\d\mu]}\right)^2
\]
If we now assume that $\d\mu\!\sim\!x\times1$Db and simply by analogy that
$\d[\d\mu]\!\sim\!x^2\times1$Db, we have
$3\Theta_4|\mu'|^2\!\sim\! 2\cdot10^3|\mu''|^2$ and thus a negligible effect
due to $\D P_{(harm.)}$.
However, it should be borne in mind that apart from the necessity of
estimating $\d[\d\mu]$ also the function $C'(\om_i)$ occuring in the
expression for $\D P_{(harm.)}$ is unknown.
On the other hand, an experimental determination of the relative relevance of
$\D P$ and $\D P_{(harm.)}$ appears feasible due to the fact that the effects
are of opposite sign.
\section*{IV. Conclusions}
We have calculated the response of a system consisting of a collection of
Brownian oscillators to the NHB field sequence.
For simplicity, we have restricted ourselves to a classical calculation and
assumed Ohmic friction, i.e. time-independent damping constants.
A more general case will be treated in a separate publication\cite{uli.ou}.
Furthermore, we have restricted the calculations to a quadratic expansion of
the dipole moment in the normal-mode coordinates $q_i$ and included no higher
terms than the quartic anharmonicity ($\propto\Theta_4$) in the potential.
We did not allow for any coupling among the modes.
We do not expect that any of these assumptions presents a severe restriction
on the applicability of our results.
Additionally, we expect a similar behavior in case that the light couples
predominantly to the polarizability instead of the dipole moment.

We have applied our calculation to the situation encountered in amorphous
systems in the frequency range of the so-called boson peak.
We do not claim that a BO-model quantitatively describes the features of the
vibrational excitations in this frequency regime. However, there appears to be
some consensus about the quasi-harmonic nature of these vibrations.
Although different models for the dynamics in the boson peak regime attribute
different damping mechanisms\cite{GKK89, SDG98, novikov98}, experimental
results often can be fitted to a simple damped oscillator susceptibility
function\cite{IXS98}.
Because less is known about the detailed form of the DOS in the THz range,
we simply parameterized it as $g(\om)\!\propto\!\om^m$. In contrast to the
situation of low-frequency Raman scattering, we are concerned with a nonlinear
response and therefore have to face four-point spatial correlation functions
of some elasto-optical constants. In a simple mean-field treatment we
factorized these four-point correlation functions.
For the various contributions to the nonlinear response different two-point
correlation functions are relevant.
We have focussed predominantly on the contribution that depends on the square
of the so-called light to vibration coupling.
For the latter, we used a functional form $C(\om)\!=\!\om^n$ with various
values for $n$ in order to be as flexible as possible and not to rely on some
specific model.
Along with the behavior of the DOS we thus can investigate the behavior for
models ranging from Debye-like phonons\cite{MB74} to soft-potential
quasi-harmonic modes\cite{GKK89}.

Our main finding is that also in the regime of oscillatory motions is NHB
able to 'detect' dynamic heterogeneities via a pump-frequency selective
modification of the linear response.
In contrast to slow relaxation processes, where usually the response to a
step field is monitored in the time domain it seems to be advantageous to
consider the fourier-transform of the response to a field pulse in case of
underdamped motion.
For all relevant pairs ($m$,$n$) we find a linear relation between the
frequency of the maximum modification and the pump-frequency $\Om$.
For underdamped modes oscillators with $\om_i\!\sim\!\Om$ are addressed
primarily whereas in the overdamped case those with $\L_i\!\sim\!\Om$ are
most relevant.
The latter situation is similar to what has been found in earlier
investigations.
If the dynamic range of the experiment is large enough, a cross-over from
underdamped oscillations to overdamped relaxation should be observable in
principle.

An important point in the context of nonlinear response functions regards the
relation between the pulse- and the step-response.
Whereas these functions are trivially related in the linear regime this
has been shown not to be the case for the nonlinear modifications. We
explicitly demonstrated this behavior for the strongly overdamped case,
i.e. the OU limit.
For this case a modification of the step-response has been found which is
very similar to the expression given earlier for a model of stochastic
dipole reorientations\cite{epl.nhb}.

Regarding the magnitude of the expected nonlinear effects, we estimate
that pump-field amplitudes on the order of $10^2\cdots10^4$kV/cm should
suffice to find nonlinear modifications which are on the order of one percent
of the linear response of the system.
With the newest technical achievements, this leads us to conclude that the
experimental realization of NHB and other nonlinear experiments in the THz
regime should be practible.

In conclusion, we have shown that NHB experiments in the THz region are
expected to yield valuable information about the physical origin of the
long-discussed vibrations around the boson peak.
In particular the question as to which extent the dynamics in this regime
has to be viewed as dynamic heterogeneous can be answered, thus shedding
further light on the nature of the dynamics in disordered systems.

\vspace{0.2cm}
\subsection*{Acknowledgements:}
We are grateful to R. Schilling and K.A. Nelson for fruitful discussions
on the topic and the reading of a first version of the manuscript.
This work was supported by the DFG under Contract No. Di693/1-1.
\newpage
\begin{appendix}
\section*{Appendix}
\setcounter{equation}{0}
\renewcommand{\theequation}{A.\arabic{equation}}
In this Appendix we give the explicit expressions for the modification of
the response, $\D R_i^{\rm(X)}(t,t_w,t_p)$, that are used in the text.

Using the abbreviation
\[
\xi_{a;b}(t)=\xi_{b;a}(t)={1\over a-b}\left(e^{-at}-e^{-bt}\right)
\]
the functions $h^{\rm X}(a,b;t)$ and $g^{\rm X}(a,b;t)$ are given by:
\Be\label{h.g.p.def}
&&h^{\rm P}(a,b;t)={1\over b}\left[\xi_{3b;a}(t)-\xi_{a+2b;b}(t)\right]
\nonumber\\
&&g^{\rm P}(a,b;t)={2\over a+b}\left[\xi_{a+2b;a}(t)-\xi_{b+2a;b}(t)\right]
\Ee
\Be\label{h.g.s.def}
&&h^{\rm S}(a,b;t)={1\over b^2}\left[\xi_{2b;a}(t)+\xi_{a+2b;2b}(t)
			    -\xi_{3b;a}(t)-\xi_{a+2b;b}(t)\right]
\nonumber\\
&&g^{\rm S}(a,b;t)={2\over a+b}\left[\xi^2_{a+b/2;b/2}(t)-\xi^2_{b+a/2;a/2}(t)
				       \right]
\Ee
and
\Be\label{h.g.ac.def}
&&h^{\rm AC}(a,b;t)=\frac{1}{b}\left[
			 \frac{1}{a+b}
			  \left(
			   \xi_{2b+i\omega,2b+a}(t)-\xi_{2b+i\omega,b}(t)
			  \right)\right.
			  \nonumber\\
		      &&\hspace{3.0cm}\left.+\frac{1}{3b-a}
			  \left(\xi_{2b+i\omega,a}(t)-\xi_{2b+i\omega,3b}(t)
			  \right)
			\right]
\nonumber\\
&&g^{\rm AC}(a,b;t)={2\over a+b}\left[
			{1\over a}
			  \left(
			    \xi_{a+b+i\omega,2a+b }(t)-\xi_{a+b+i\omega,b}(t)
			    \right)\right.\\
		      &&\hspace{3.6cm}\left.-{1\over b}
			  \left(
			   \xi_{a+b+i\omega,2b+a }(t)-\xi_{a+b+i\omega,a}(t)
			  \right)
			 \right]\nonumber
\Ee

In order to calculate the fourier transform of $R_i^{\rm(P)}(t)$ and
$\D R_i^{\rm(P)}(t,t_w,t_p)$ it is sufficient to note that these can
easily be obtained from eqns.(\ref{Ri.p.t}) and (\ref{del.Ri.X}) by noting
that
\[
{\cal F}[\xi_{a;b}(t)]={1\over a-b}
\left({1\over a-i\om}-{1\over b-i\om}\right).
\]
\end{appendix}

\newpage
\subsection*{Figure captions}
\begin{description}
\item[Fig.1 : ] The field sequence for the nonresonant hole burning (NHB)
experiment: One or more cycles of a strong sinusoidal field
$E_P(t)=E_P\sin{(\Om t)}$ are applied to a sample in thermal equilibrium.
After a waiting time $t_w$ the response to a small field $E_M(t)$ is
monitored.
\item[Fig.2 : ] The dielectric loss $\e''(\om)$ according to eq.(\ref{eps.om})
versus frequency for $g(\om_i)=(3/\om_c^3)\om_i^2\theta(\om_c-\om_i)$,
$C(\om_i)=\om_i$ and $\rho/(m\e_0)|\mu'|^2=1$.
The damping constant $\g$ is chosen as $\g\!=\!0,0.01,0.1,1.0,100$ and
$\om_c\!=\!10$.
from bottom to top. Also shown is $\e_D''(\om)$ for a Debye relaxation (thin
dotted line) for comparison.
Note that for $\g\!=\!100$ the motion is strongly overdamped in the whole
frequency range.
\item[Fig.3 : ] Imaginary (upper panel) and real (lower panel) part of
$\D R^{\rm(P)}_i(\om,t_w,t_p)$ (abbreviated as $\D R^{\rm(P)}_i(\om)$)
versus measuring frequency $\om$ for $\om_i=1.0,\g_i=0.2$ (full lines;
underdamped case) and $\om_i=10.0,\g_i=100.0$ (dashed lines; overdamped case).
One cycle of the sinusoidal pump-field with frequency $\Om=1.0$ has been used
and the waiting was set to zero, $t_w=0$.
$\D R^{\rm(P)}_i(\om,t_w,t_p)$ is obtained from
$\D R^{\rm(P)}_i(t,t_w,t_p)$, eq.(\ref{del.Ri.X}), via fourier transform as
explained in the text.
The results for the overdamped case are multiplied by a factor $10^8$.
\item[Fig.4 : ]{\bf a}: $\D P^{\rm(P)}(\om)''$ versus $\om$ for
various burn frequencies $\Om$ (upper panel). The DOS and the light to
vibration coupling constant are the same as in Fig.2. The other parameters are
$(3\rho/m^3\Theta_4E_ME_P^2|\mu'|^4)=1$, $t_w=0$ and $\g=10^{-2}\om_c$.
The burn frequencies are chosen as $\Om/\om_c=0.0003,0.001,0.003$
(dashed lines) and $\Om/\om_c=0.01,0.03,0.1,0.3$ (full lines).
The dashed lines present overdamped oscillators while the full lines
correspond to underdamped modes.
In the lower panel $\D P_{(harm.)}^{\rm(P)}(\om)''$ is plotted, where
additionally $C'(\om_i)\!=\!C(\om_i)$ is assumed and we set
$(\rho/m^3E_ME_P^2|\mu'|^2|\mu''|^2)=1$.\\
{\bf b}: Upper panel: The frequency at which $\D P^{\rm(P)}(\om)''$
takes its minimum value (cf. Fig.4a), $\om_{\rm min}/\om_c$ versus burn
frequency $\Om$. Lower panel: $\om_{\rm max}/\om_c$ versus $\Om$ for
$\D P_{(harm.)}^{\rm(P)}(\om)''$.
The parameters for the DOS and the light to vibration coupling constants are
written as $g(\om)\propto\om^m$ and $C(\om)=\om^n$. The results shown are for
two pairs ($m$,$n$) and a damping constant $\g=10^{-2}\om_c$.\\
\item[Fig.5 : ] $\D R^{\rm(P)}(t,t_w,t_p)$ (upper panel) and
$\D R_{(harm.)}^{\rm(P)}(t,t_w,t_p)$ (lower panel) versus measuring time for
the same parameters as used in Fig.4, $\Om/\om_c\!=\!0.1$ and various numbers
of pump-field cycles, $N$.
The dotted envelopes are of the form $te^{-\g t}$.
In the lower panel, the envelope has been divided by a factor of $30$.
\end{description}
\newpage
\begin{figure}
\includegraphics[width=15cm]{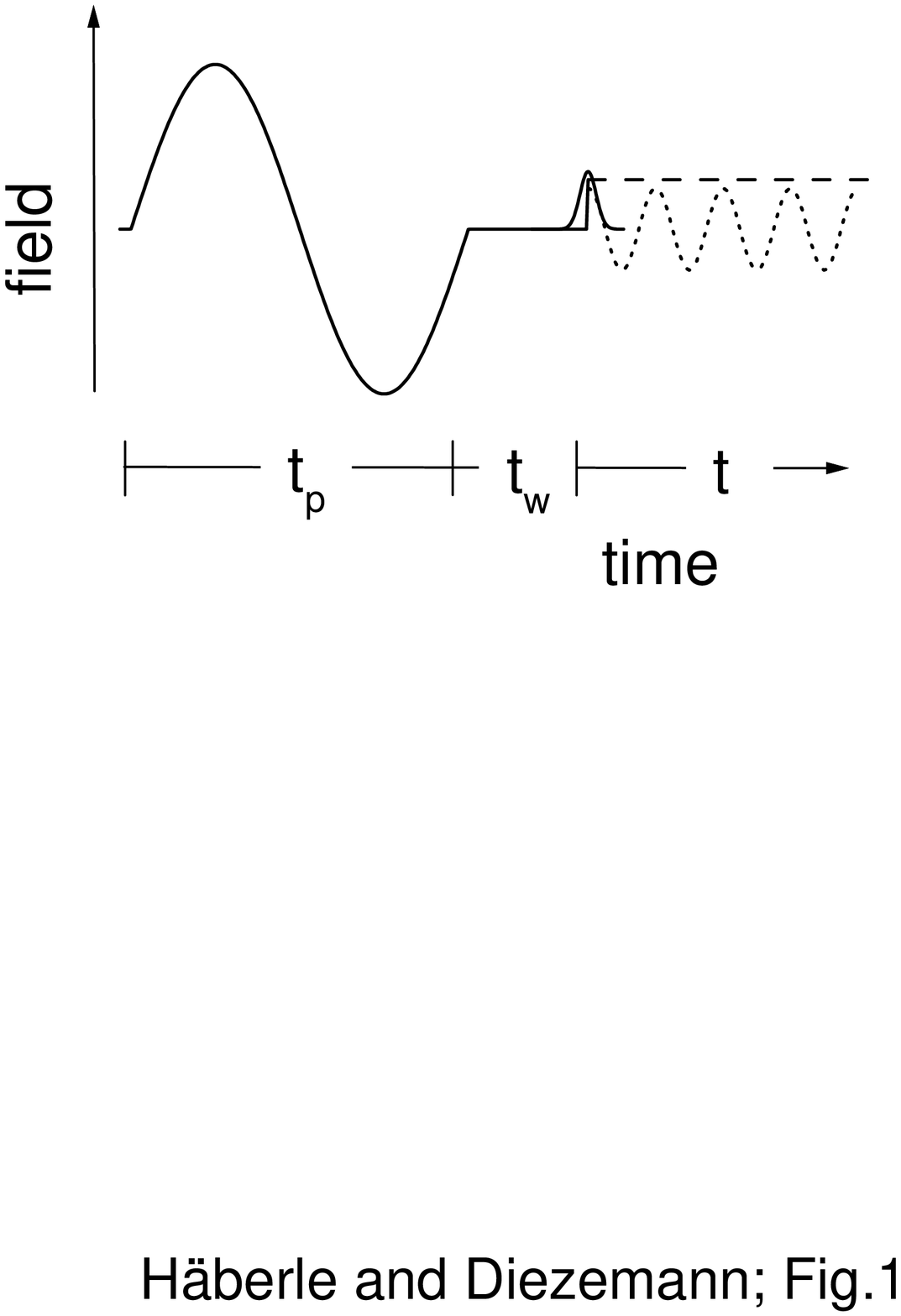}
\end{figure}
\newpage
\begin{figure}
\includegraphics[width=15cm]{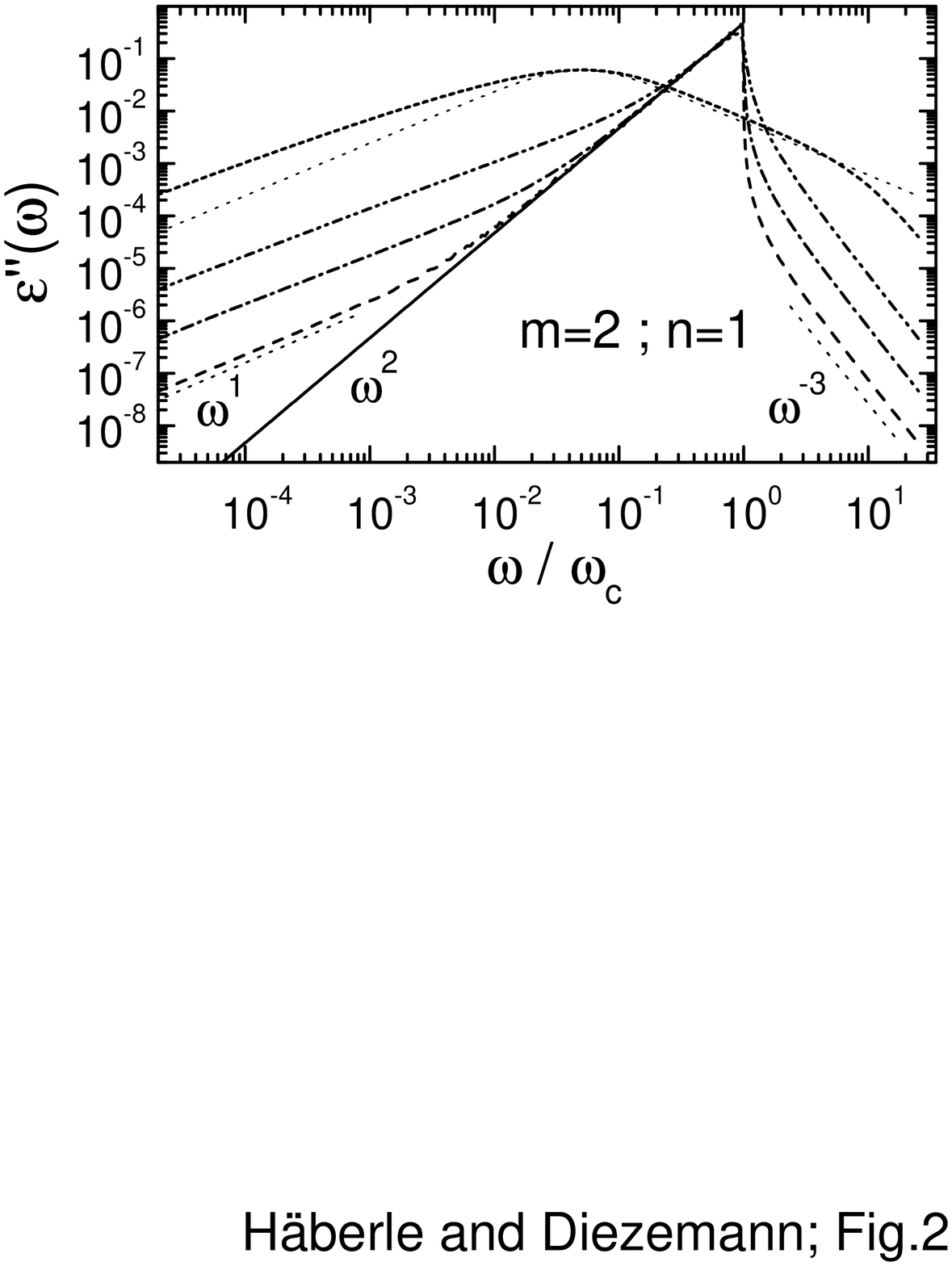}
\end{figure}
\newpage
\begin{figure}
\includegraphics[width=15cm]{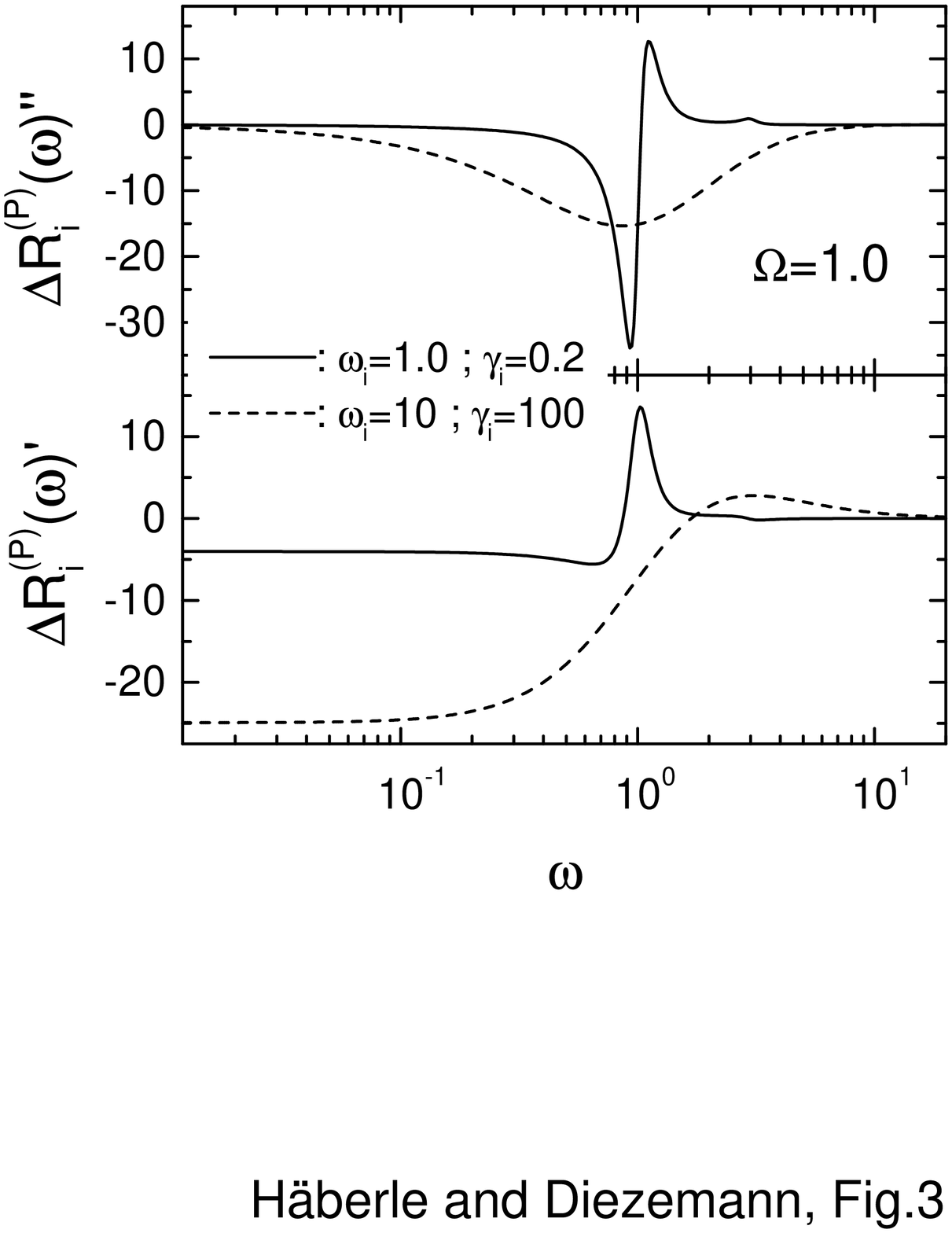}
\end{figure}
\newpage
\begin{figure}
\includegraphics[width=15cm]{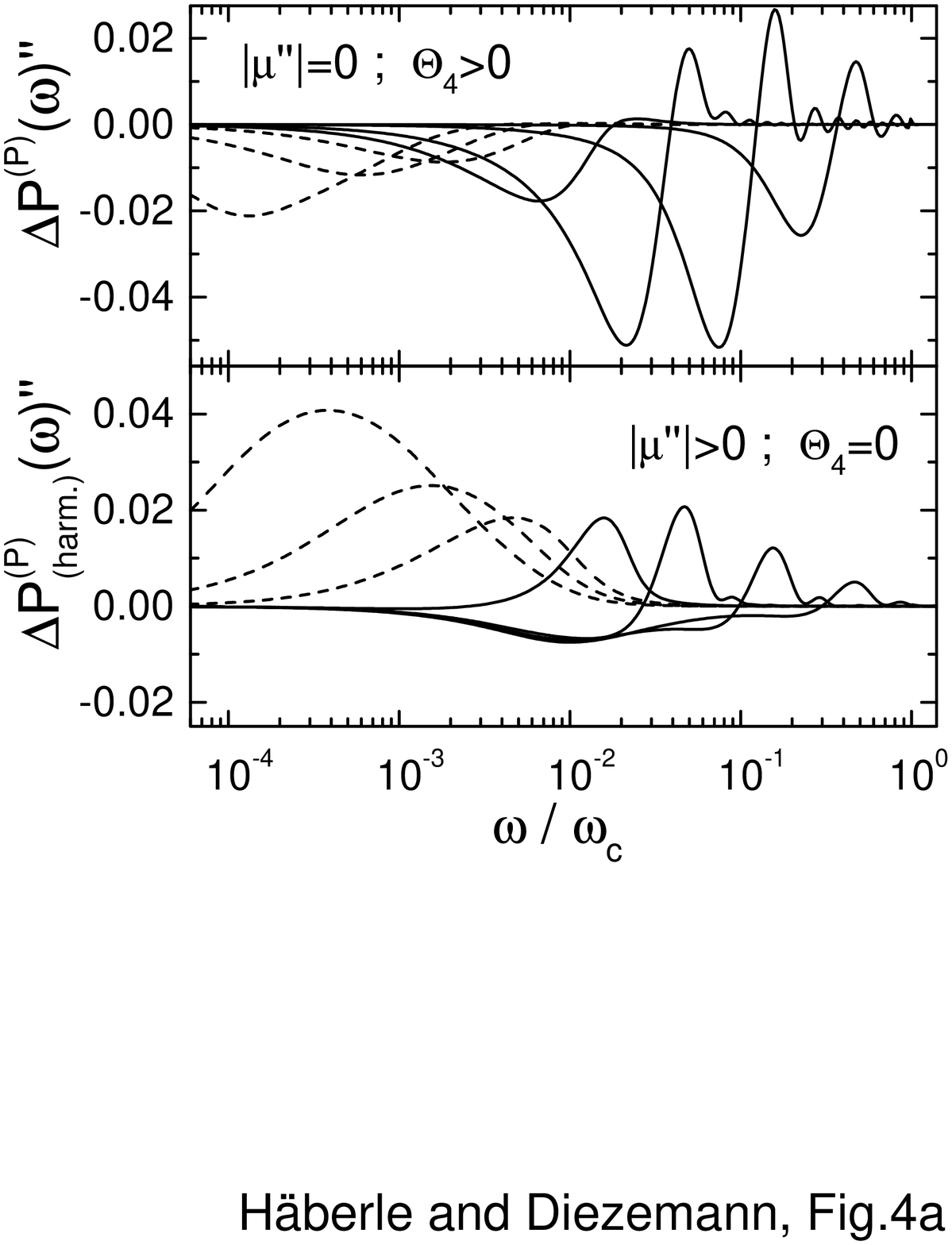}
\end{figure}
\newpage
\begin{figure}
\includegraphics[width=15cm]{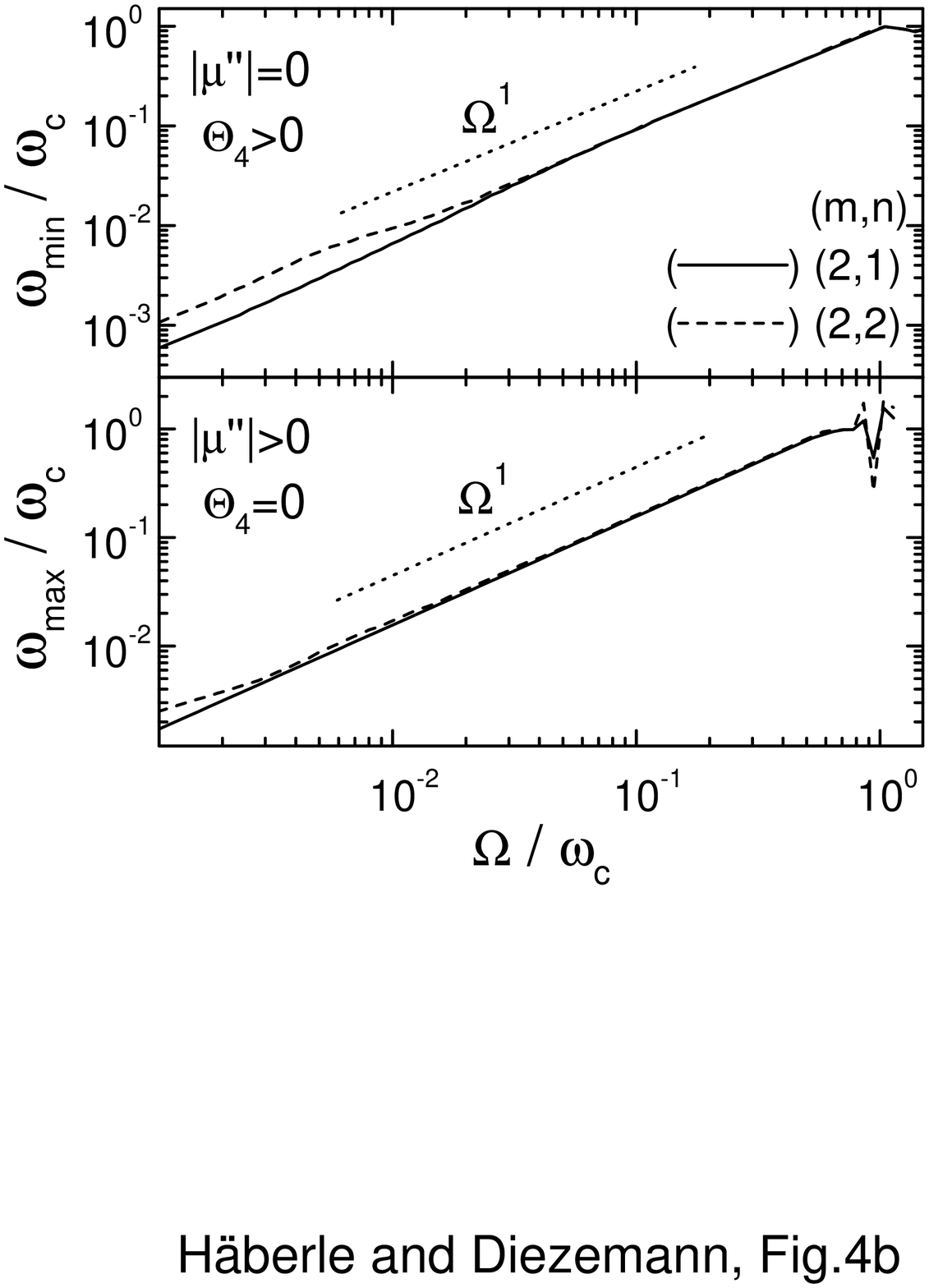}
\end{figure}
\newpage
\begin{figure}
\includegraphics[width=15cm]{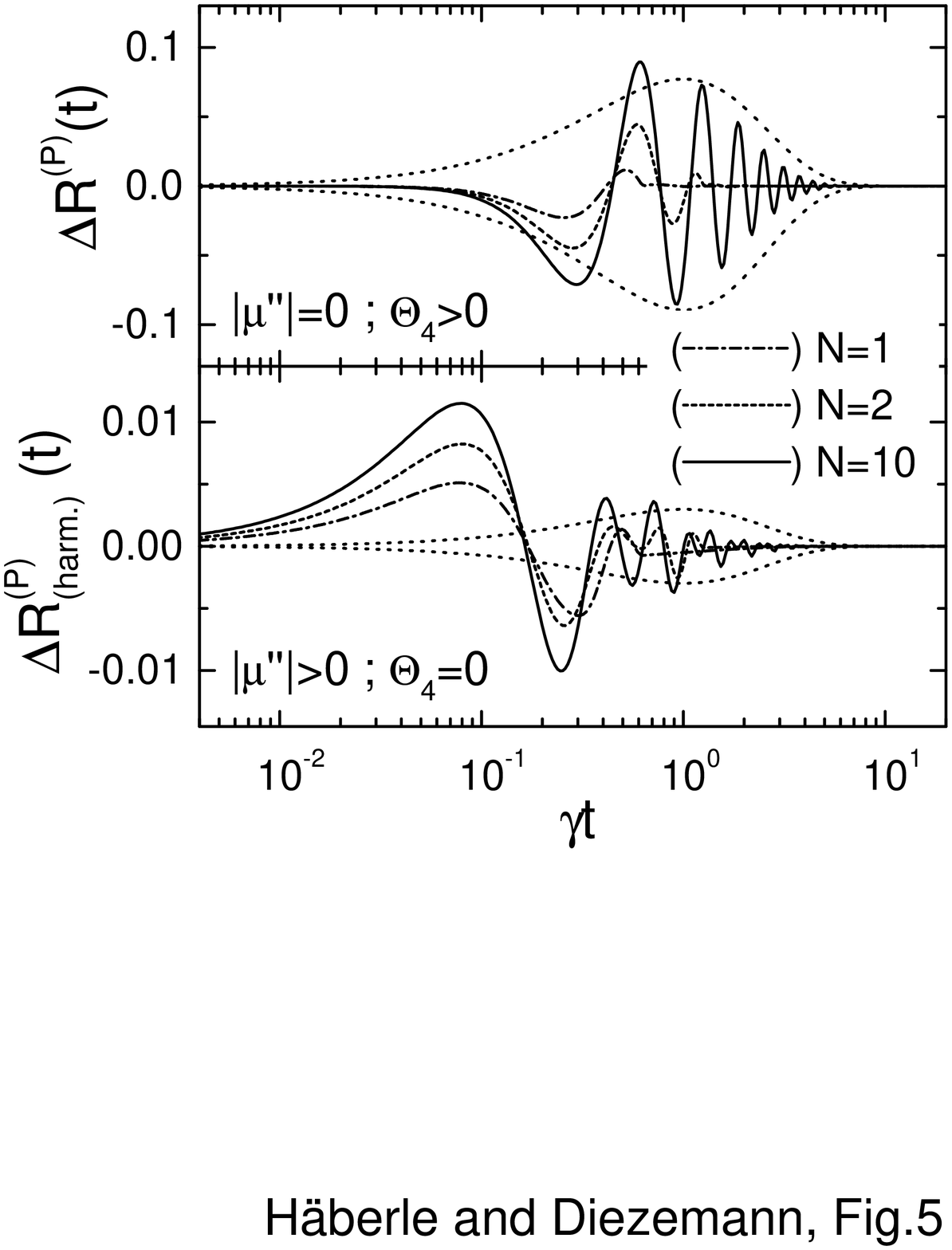}
\end{figure}
\end{document}